\documentclass[12pt,cite,epsf,epsfig]{article}
\usepackage{epsfig}
\usepackage{float}
\usepackage{cite}
\usepackage{amsmath}

\begin{document}
\author{S. Dev \thanks{sdev@associates.iucaa.in} $^{,1}$,
Desh Raj \thanks{raj.physics88@gmail.com} $^{,2}$, Radha Raman Gautam \thanks{gautamrrg@gmail.com} $^{,2}$}
\date{$^1$\textit{Department of Physics, School of Sciences, HNBG Central University, Srinagar, Uttarakhand 246174, INDIA.}\\
\smallskip
$^2$\textit{Department of Physics, Himachal Pradesh University, Shimla 171005, INDIA.}}
\title{Deviations in Tribimaximal Mixing From Sterile Neutrino Sector}
\maketitle
\begin{abstract}
We explore the possibility of generating a non-zero $U_{e3}$ element of the neutrino mixing matrix from tribimaximal neutrino mixing by adding a light sterile neutrino to the active neutrinos. Small active-sterile mixing can provide the necessary deviation from tribimaximal mixing to generate a non-zero $\theta_{13}$ and atmospheric mixing $\theta_{23}$ different from maximal. Assuming no CP-violation, we study the phenomenological impact of sterile neutrinos in the context of current neutrino oscillation data. The tribimaximal pattern is broken in such a manner that the second column of tribimaximal mixing remains intact in the neutrino mixing matrix.
\end{abstract}
\section{Introduction}
With the advent of precision neutrino measurements, the focus has shifted to the determination of the unknown parameters such as the neutrino mass ordering, the leptonic CP violation and the absolute neutrino mass scale. On the other hand, Beyond the Standard Model (BSM) physics scenarios such as non-standard neutrino interactions, unitarity violation, CPT- and Lorentz-invariance violation and models with sterile neutrinos are being investigated vigorously. Several anomalies at short baselines hint towards the existence of one or more sterile neutrinos at eV scale or even higher. The evidence for $\overline{\nu}{_{\mu}\rightarrow\overline{\nu}_{e}}$ appearance in the LSND experiment \cite{lsnd} subsequently confirmed by the MiniBooNE experiment \cite{miniboone} in both the neutrino and the antineutrino modes is compatible with one or more extra sterile neutrinos at the eV scale. In addition, recent estimates of the reactor $\overline{\nu}_{e}$  fluxes \cite{reactor} strongly indicate the oscillations of electronic neutrinos into sterile neutrinos. Each of these observations may be explained by the addition of at least one extra sterile neutrino. The hints for the presence of sterile neutrinos come not only from neutrino physics but also from Big Bang nucleosynthesis (BBN) and the structure of the universe. There are mild hints for extra radiation in the universe in addition to the photons and active neutrinos from precision cosmology. It could, in principle, be any relativistic degree of freedom. In fact, most of the recent cosmological parameter fits are compatible with more radiation than the Standard Model (SM) particle content. Further support for the presence of extra radiation comes from the higher $^{4}H_{e}$  abundance \cite{he}. However, the most recent data from Planck \cite{planck} strongly disfavour fully thermalized neutrinos  with mass $\approx$ 1 eV which have been proposed to explain the neutrino anomalies at short baselines. However, the latest Planck data \cite{planck} does not exclude the possibility of heavier $(\geq$ 1 eV) sterile neutrinos. Planck limits the effective number of relativistic degrees of freedom to $N_{eff}=3.15\pm0.23$ and the sum of neutrino masses $\sum m_{\nu}\leq 0.23 eV$ which is in good agreement with the standard model of cosmology with $N_{eff} = 3.046$. Therefore, adding light sterile species would result in tension with the cosmological bounds. This conflict can be resolved if eV scale sterile neutrinos are partially thermalized before BBN era but equilibrate with active neutrinos at a later time. This can happen if sterile neutrinos have self-interactions. The self interactions can induce large matter potential at high temperatures, suppress the effective mixing angle and block production of sterile neutrinos from oscillations. As the universe cools down, flavor equilibrium between active and sterile species can be reached after BBN epoch which leads to a decrease of $N_{eff}$. The conflict with cosmological neutrino mass bounds on the additional sterile neutrinos can be relaxed if more light sterile species are introduced. Complete analysis is given in reference \cite{tang}.\\
From a theoretical standpoint, sterile neutrinos are a natural consequence of non-zero neutrino mass. Sterile neutrinos are SM singlets and are, as such, subject to gravitational interactions only. Since they do not interact weakly like active neutrinos, sterile neutrinos are far more elusive than active neutrinos. However, sterile neutrinos could mix with active neutrinos  signalling BSM physics. Sterile neutrinos, if they indeed exist, could be produced in the early universe and may have played an important role in the cosmological evolution. The missing entities in the SM are the right-handed neutrinos which are the obvious sterile neutrino candidates. Their existence would imply left-right symmetry as well as quark-lepton symmetry which underly left-right symmetric and grand unified models, respectively. There is no compelling theoretical motivation for these gauge singlets to have small masses. In fact, the most popular models of  neutrino mass generation, the so called seesaw mechanisms for light active neutrinos require the right-handed neutrinos to be very massive Majorana fermions with a mass scale of the order of the grand unified scale. The hypotheses of the existence of light sterile neutrinos with eV scale masses and the indicated charged current couplings to electron and muon will be tested in a number of experiments with reactor and accelerator neutrinos \cite{kn}. 
The upper limit on the sterile neutrino parameters by Super-Kamiokande have constrained $\vert U_{\mu4} \vert^2 <0.041$ and $\vert U_{\tau 4}\vert^2< 0.18$ at 90\% confidence level (CL) \cite{skc}. More stringent constraints on sterile neutrinos are expected from the ongoing Daya Bay \cite{dayabay}  and upcoming JUNO \cite{juno} experiments. There are many experiments which confirm physics beyond the SM in the neutrino sector \cite{exp}. Refs. \cite{kn,kang,leonard,kyu,ivan} provide detailed analysis of experimental results in the context of light sterile neutrinos and their effect on active neutrino parameters, $N_{eff}$, cosmology, and dark matter. There are many models \cite{daijiro,rode,maria,zang,common,hong,diana,debashish,sterilemodel} discussed in the literature which give mixing of sterile neutrinos with active ones. Harrison, Perkins and Scott first showed that experimentally obtained mixing matrix is close to the so-called tribimaximal (TBM) mixing \cite{hps}. There are a plethora of neutrino mixing models derived from discrete non-Abelian symmetries \cite{tbm} leading to  TBM mixing matrix. These models with TBM mixing were quite successful in explaining experimental data until the results from various experiments \cite{th13} confirmed a sizeable non-zero reactor mixing angle ($\theta_{13}$) as a result of which the TBM based models came under intense theoretical scrutiny. As a consequence, many models \cite{tbmd} which use perturbations to modify the TBM mixing to generate non-zero $\theta_{13}$ were proposed. There, also, exist models \cite{carl,sanjeev,dev} which fix one or more columns (rows) of TBM and perturb others to generate mixing angles within their experimental ranges. A non-zero $\theta_{13}$ can also be obtained by introducing light sterile neutrinos \cite{common, diana, debashish}. In the present work, we attempt to incorporate sterile neutrinos along with active neutrinos to generate deviations from the TBM mixing while keeping one of the columns of TBM fixed. With four neutrinos mixing with each other, it is quite a formidable task to make any predictions for various neutrino parameters so we impose the additional constraint of CP-conservation to simplify the analysis. 
The present work allows non-trivial mixing between active and sterile neutrinos which are found to have interesting consequences.
The global fits to data from short baseline (SBL) neutrino experiments suggest that the data can be described by either (3+1) or (3+2) schemes with one or two sterile neutrinos, respectively. In the present work, we focus on the simplest extension viz. the (3+1) scheme with one sterile neutrino and attempt to construct the mixing matrix of (3+1) neutrinos keeping one of the columns identical to that of TBM.
\section{Methodology}
We consider a light (eV scale) sterile neutrino in addition to the three active neutrinos and set CP violating phases to be equal to zero to make the analysis simple. There are in total 10 physical parameters viz. 4 neutrino masses, 3 active mixing angles and 3 active-sterile mixing angles.\\
We define the mass matrix for $3+1$ scheme as
\begin{small} 
\begin{equation}
M_{4\times 4}=\left(%
\begin{array}{cc}
  M_{TBM} & A \\
  A^T & \bar{m}_{s} \\ 
\end{array}%
\right)
\end{equation}
\end{small}
such that the upper 3$\times$3 sector $M_{TBM}$ is diagonalized by TBM mixing matrix and the column $A$ has 3 elements belonging to the sterile sector. Specific structures of this column could have interesting consequences some of which have been discussed in Ref.\cite{common}.\\
Therefore, the mass matrix has the following form
\begin{small}
\begin{equation}
M_\nu = \left(
\begin{array}{cccc}
 \frac{1}{3} (2 \bar{m}_{1}+ \bar{m}_{2}) & \frac{1}{3}(\bar{m}_{2}-\bar{m}_{1}) & \frac{1}{3}(\bar{m}_{2}-\bar{m}_{1}) & e \\
 \frac{1}{3}(\bar{m}_{2}-\bar{m}_{1}) & \frac{1}{6} (\bar{m}_{1}+2 \bar{m}_{2}+3 \bar{m}_{3}) & \frac{1}{6} (\bar{m}_{1}+2 \bar{m}_{2}-3 \bar{m}_{3}) & f \\
 \frac{1}{3}(\bar{m}_{2}-\bar{m}_{1}) & \frac{1}{6} (\bar{m}_{1}+2 \bar{m}_{2}-3 \bar{m}_{3}) & \frac{1}{6} (\bar{m}_{1}+2 \bar{m}_{2}+3 \bar{m}_{3}) & g \\
 e & f & g & \bar{m}_{s}
\end{array}
\right),
\end{equation}
\end{small}
where $\bar{m}_{1}$, $\bar{m}_{2}$ and $\bar{m}_{3}$ are the mass eigenvalues of $3\times3$ active neutrino mass matrix.
In the(3+1) scheme, there are four massive neutrinos and the corresponding neutrino mixing matrix is a $4\times4$ unitary matrix. We use the following parametrization \cite{rode} for the mixing matrix with CP violating phases taken to be zero
\begin{small}
\begin{equation}
U_{4\times 4}= R(\theta_{34}) R(\theta_{24}) R(\theta_{14}) R(\theta_{23}) R(\theta_{13}) R(\theta_{12})
\end{equation}
\end{small}
where $R(\theta_{ij})$ matrix describes rotation in $ij^{\textrm{th}}$ plane. In this parametrization, we have
\begin{small}
\begin{eqnarray}
U_{e1}&=&\cos\theta_{12} \cos\theta_{13} \cos\theta_{14},\nonumber \\
U_{e2}&=&\cos\theta_{14} \cos\theta_{13} \sin\theta_{12},\nonumber \\
U_{e3}&=& \cos\theta_{14} \sin\theta_{13},\nonumber\\
U_{e4}&=& \sin\theta_{14}, \\
U_{\mu4}&=&\cos\theta_{14} \sin\theta_{24},\nonumber\\
U_{\tau4}&=&\cos\theta_{14} \cos\theta_{24} \sin\theta_{34}, \nonumber\\
U_{s4}&=&\cos\theta_{14} \cos\theta_{24} \cos\theta_{34}. \nonumber\\
U_{\mu3}&=& \cos\theta_{13} \cos\theta_{24}\sin\theta_{23}-\sin\theta_{13} \sin\theta_{14} \sin\theta_{24}.\nonumber
\end{eqnarray}
\end{small}
Since CP violation is neglected in our analysis, the neutrino mass matrix is real and the columns of mixing matrix are given by normalized eigenvectors of the mass matrix $M_{\nu}$. The $3\times3$ active neutrino sector of $M_{\nu}$ is still diagonalized by TBM mixing matrix, we only need three rotation matrices along with TBM to completely diagonalize the neutrino mass matrix $M_{\nu}$. Since we are interested in the cases where one of the columns of TBM remains intact in the final mixing matrix, the resulting mixing matrix is somewhat similar to the TM$_{1}$/TM$_{2}$ variants of TBM \cite{carl,sanjeev,dev}. In the present work, TM$_{1}$/TM$_{2}$ are $4\times4$ neutrino mixing matrices having the first/second column same as that of TBM. The third column of the neutrino mixing matrix cannot be the same as that of TBM as this gives $|U_{e3}|=0$, which is inconsistent with the  current experimental data. The TM$_1$ form of the mixing matrix is obtained when we substitute $e = \frac{f+g}{2}$ in the mass matrix given in Eq. (2). We find that TM$_{1}$ mixing is phenomenologically ruled out for the CP-conserving case because the contribution from sterile sector cannot simultaneously keep $\theta_{13}$ and $\theta_{23}$ within their current experimentally allowed ranges. The only viable case is TM$_{2}$ in which the second column of mixing matrix is the same as that of TBM.\\
If we substitute $e = -(f + g)$ in the mass matrix in Eq. (2), the resulting mass matrix is of the TM$_2$ type. The mass matrix is modified to the following form which gives the mixing matrix of TM$_{2}$ type:
\begin{small}
\begin{equation}
M_\nu=\left(
\begin{array}{cccc}
 \frac{1}{3} (2 \bar{m}_{1}+ \bar{m}_{2}) & \frac{1}{3}(\bar{m}_{2}-\bar{m}_{1}) & \frac{1}{3}(\bar{m}_{2}-\bar{m}_{1}) & -(f+g) \\
 \frac{1}{3}(\bar{m}_{2}-\bar{m}_{1}) & \frac{1}{6} (\bar{m}_{1}+2 \bar{m}_{2}+3 \bar{m}_{3}) & \frac{1}{6} (\bar{m}_{1}+2 \bar{m}_{2}-3 \bar{m}_{3}) & f \\
 \frac{1}{3}(\bar{m}_{2}-\bar{m}_{1}) & \frac{1}{6} (\bar{m}_{1}+2 \bar{m}_{2}-3 \bar{m}_{3}) & \frac{1}{6} (\bar{m}_{1}+2 \bar{m}_{2}+3 \bar{m}_{3}) & g \\
 -(f+g) & f & g & \bar{m}_{s}
\end{array}
\right).
\end{equation}
\end{small}
The modified mass matrix $M_\nu$ can be diagonalized as
\begin{equation}
M_{dig}=U^{T}_{\nu} M_{\nu} U_{\nu}
\end{equation}
where $U_{\nu}= U_{TBM}~R(\bar{\theta}_{34})~R(\bar{\theta}_{14})~R(\bar{\theta}_{13})$.\\
The mixing matrix $U_{\nu}$ takes the following form
\begin{small}
\begin{equation}
U_{\nu}=\left(
\begin{array}{cccc}
 \sqrt{\frac{2}{3}} \bar{c}_{14} \bar{c}_{13} & \frac{1}{\sqrt{3}} & \sqrt{\frac{2}{3}} \bar{c}_{14} \bar{s}_{13} & \sqrt{\frac{2}{3}} \bar{s}_{14} \\
 \frac{\bar{c}_{34} \bar{s}_{13} +\bar{c}_{13} \bar{s}_{14} \bar{s}_{34}
  }{\sqrt{2}}-\frac{\bar{c}_{14} \bar{c}_{13}}{\sqrt{6}} & \frac{1}{\sqrt{3}} & -\frac{3 
  \bar{c}_{13} \bar{c}_{34} +\bar{s}_{13} \left(\sqrt{3} \bar{c}_{14} -3 \bar{s}_{14} \bar{s}_{34}
  \right)}{3 \sqrt{2}} & -\frac{\bar{s}_{14}}{\sqrt{6}}-\frac{\bar{c}_{14} \bar{s}_{34}
  }{\sqrt{2}} \\
 -\frac{\sqrt{3} \bar{c}_{14} \bar{c}_{13} +3 \bar{s}_{14} \bar{s}_{34} \bar{c}_{13} +3 
   \bar{c}_{34} \bar{s}_{13} }{3 \sqrt{2}} & \frac{1}{\sqrt{3}} & \frac{3 \bar{c}_{13} \bar{c}_{34}
   -\bar{s}_{13} \left(\sqrt{3} \bar{c}_{14} +3 \bar{s}_{14} \bar{s}_{34} \right)}{3 \sqrt{2}}
   & \frac{\bar{c}_{14} \bar{s}_{34}}{\sqrt{2}}-\frac{\bar{s}_{14}}{\sqrt{6}} \\
 \bar{s}_{13} \bar{s}_{34} -\bar{c}_{13} \bar{c}_{34} \bar{s}_{14} & 0 & -\bar{c}_{34} \bar{s}_{14} \bar{s}_{13} -\bar{c}_{13} \bar{s}_{34} & \bar{c}_{14} \bar{c}_{34}
\end{array}
\right)
\end{equation}
\end{small}
where $\bar{c}_{ij}=\cos\bar{\theta}_{ij}$ and $\bar{s}_{ij}=\sin\bar{\theta}_{ij}$.\\
Rotation angles $\bar{\theta}_{14}$, $\bar{\theta}_{13}$, $\bar{\theta}_{34}$ and the mass matrix elements $f, g, \bar{m}_{s}$ are related as
\begin{small}
\begin{eqnarray}
f&=&\frac{8 \bar{m}_{1} \sin \bar{\theta}_{14} q+\bar{m}_{3} r+\bar{m}_{s} \cos \bar{\theta}_{14} p} 
   {8 \sqrt{6} \left(\cos ^2\bar{\theta}_{14} \cos 2 \bar{\theta}_{34} -\sin ^2\bar{\theta}_{14} \right) \left(\cos \bar{\theta}_{14} \cos 2 \bar{\theta}_{34} -\sqrt{3} \sin \bar{\theta}_{14} \sin \bar{\theta}_{34} \right)},\nonumber\\
g&=&\frac{2 \bar{m}_{1} w+\bar{m}_{3} \tan \bar{\theta}_{34} v+4 \bar{m}_{s} \cos \bar{\theta}_{14} \cos \bar{\theta}_{34} u}{3 \sqrt{2} \left(4 \cos ^2\bar{\theta}_{14} \cos 2 \bar{\theta}_{34} -4 \sin ^2\bar{\theta}_{14} \right)},\\
\bar{m}_{s}&=&\frac{2 \bar{m}_{1} y+\bar{m}_{3} x}
{\sin 2
   \bar{\theta}_{13} \left(2 \cos 2 \bar{\theta}_{14} \cos ^2\bar{\theta}_{34} -3 \cos 2 \bar{\theta}_{34} +1\right)-4 \sin \bar{\theta}_{14}
   \cos 2 \bar{\theta}_{13} \sin 2 \bar{\theta}_{34} },\nonumber
\end{eqnarray}
\end{small}
where
\begin{small}
\begin{eqnarray}
p&=& 4 \sqrt{3} \sin ^2 \bar{\theta}_{14} \sin 2 \bar{\theta}_{34} -2 \sqrt{3} \cos
   ^2 \bar{\theta}_{14} \sin 4 \bar{\theta}_{34} +2 \sin \bar{\theta}_{14} \cos \bar{\theta}_{14}
   (\cos \bar{\theta}_{34} -5 \cos 3 \bar{\theta}_{34}),\nonumber \\
q&=& \cos ^2 \bar{\theta}_{14} \cos ^2 2 \bar{\theta}_{34} \sec \bar{\theta}_{34} -3 \sin
   ^2 \bar{\theta}_{14} \sin \bar{\theta}_{34} \tan \bar{\theta}_{34} ,\nonumber \\
r&=& 24 \sin ^3\bar{\theta}_{14} \sin \bar{\theta}_{34} \tan \bar{\theta}_{34} +2 \sqrt{3}
   \cos ^3\bar{\theta}_{14} \sin 4 \bar{\theta}_{34} -4 \sqrt{3} \sin ^2 \bar{\theta}_{14}
   \cos \bar{\theta}_{14} \sin 2 \bar{\theta}_{34}\nonumber \\
   &&-4 \sin \bar{\theta}_{14} \cos
   ^2\bar{\theta}_{14} \sin \bar{\theta}_{34} (\cos 2 \bar{\theta}_{34} +3) \tan \bar{\theta}_{34},\nonumber \\
u&=& 3 \cos \bar{\theta}_{14} \sin \bar{\theta}_{34} -\sqrt{3} \sin \bar{\theta}_{14} , \\
v&=& -6 \cos ^2\bar{\theta}_{14} \cos 2 \bar{\theta}_{34} +4 \sqrt{3} \sin \bar{\theta}_{14}
   \cos \bar{\theta}_{14} \sin \bar{\theta}_{34} -9 \cos 2 \bar{\theta}_{14} +3,\nonumber \\
w&=& \sqrt{3} \sin 2 \bar{\theta}_{14} \cos 2 \bar{\theta}_{34} \sec \bar{\theta}_{34} -6 \sin
   ^2\bar{\theta}_{14} \tan \bar{\theta}_{34} ,\nonumber \\
x&=& -\sin 2 \bar{\theta}_{13} \left(\cos 2 \bar{\theta}_{14} (\cos 2 \bar{\theta}_{34} -5)+6
   \sin ^2\bar{\theta}_{34} \right)-8 \sin \bar{\theta}_{14} \cos 2 \bar{\theta}_{13} \sin
   ^2\bar{\theta}_{34} \tan \bar{\theta}_{34} ,\nonumber \\
y&=& \cos 2 \bar{\theta}_{34} ((\cos 2 \bar{\theta}_{14} -3) \sin 2 \bar{\theta}_{13} -4 \sin
   \bar{\theta}_{14} \cos 2 \bar{\theta}_{13} \tan \bar{\theta}_{34} )+4 \sin ^2\bar{\theta}_{14}
   \sin 2 \bar{\theta}_{13}.\nonumber
\end{eqnarray}
\end{small}
Using Eqs.(4) and (7), we obtain the six mixing angles  
\begin{small}
\begin{eqnarray}
\sin\theta_{14}&=& \sqrt{\frac{2}{3}} \sin \bar{\theta}_{14} \nonumber,\\
\sin\theta_{24}&=& \sec \theta_{14} \left|-\frac{\sin \bar{\theta}_{14} }{\sqrt{6}}-\frac{\cos \bar{\theta}_{14} \sin \bar{\theta}_{34}
  }{\sqrt{2}}\right|\nonumber,\\
\sin\theta_{34}&=& \sec \theta_{14} \sec \theta_{24} \left|\frac{\cos \bar{\theta}_{14} \sin \bar{\theta}_{34}
  }{\sqrt{2}}-\frac{\sin \bar{\theta}_{14}}{\sqrt{6}}\right|,\\
\sin\theta_{13}&=&\sqrt{\frac{2}{3}} \sec \theta_{14} |\cos \bar{\theta}_{14} \sin \bar{\theta}_{13}|\nonumber,\\
\sin\theta_{12}&=& \frac{\sec \theta_{13} \sec \theta_{14}}{\sqrt{3}}\nonumber,\\
\sin\theta_{23}&=& \left| U_{\mu3}\right| \sec\theta_{13} \sec\theta_{24}+\sin\theta_{14} \tan\theta_{13} \tan\theta_{24}.\nonumber
\end{eqnarray}
\end{small}
The neutrino masses are given by
\begin{small}
\begin{eqnarray}
m_{1} &= & \bar{m}_{1}-\frac{(\bar{m}_{1}-\bar{m}_{3}) \sin \bar{\theta}_{14} \sin \bar{\theta}_{13} \sec\bar{\theta}_{34}}{\sin \bar{\theta}_{14} \sin \bar{\theta}_{13} \cos \bar{\theta}_{34}+\cos \bar{\theta}_{13} \sin \bar{\theta}_{34}}, \nonumber \\
m_{2} & = & \bar{m}_{2}, \nonumber \\
m_{3} & = &\bar{m}_{1}-\frac{(\bar{m}_{1}-\bar{m}_{3}) \sin \bar{\theta}_{14} \cos\bar{\theta}_{13} \sec \bar{\theta}_{34}}{\sin \bar{\theta}_{14} \cos \bar{\theta}_{13} \cos \bar{\theta}_{34}-\sin \bar{\theta}_{13} \sin \bar{\theta}_{34}},\\
m_{4} & = &\bar{m}_{1}+\frac{8 (\bar{m}_{1}-\bar{m}_{s}) \cos ^2\bar{\theta}_{14} \cos \bar{\theta}_{34}}{16 \sin \bar{\theta}_{14}\cot (2 \bar{\theta}_{13} ) \sin ^3(\bar{\theta}_{34} )+\cos (2 \bar{\theta}_{14} ) (\cos (3 \bar{\theta}_{34} )-9 \cos \bar{\theta}_{34})+6 \sin (2 \bar{\theta}_{34} ) \sin \bar{\theta}_{34})}. \nonumber  
\end{eqnarray}
\end{small}
It is clear from Eqs.(11) that the active neutrino masses $m_{1}$, $m_{3}$ are modified from their original values $\bar{m}_{1}$ and $\bar{m}_{3}$ while the eigenvalue $m_{2}$, which corresponds to the second eigenvector of the mass matrix remains unchanged. 
\section{Numerical Analysis}
The presence of sterile neutrino(s) affects the active neutrino mixing angles via the unitarity conditions of the mixing matrix i.e., $\Sigma_{j}\vert U_{ij}\vert^{2}=1$, where $i=e,\mu,\tau,s$ and $j=1,2,3,4$. In our numerical analysis, we use the 3$\sigma$ ranges of the neutrino oscillation parameters \cite{kopp}. Experimental constraints on mass squared differences of active neutrino parameters at 3$\sigma$ are $\Delta m_{21}^{2} = (7.11-8.18) \times 10^{-5} eV^{2}$ and $|\Delta m_{31}^{2}| = 2.30-2.65 \times 10^{-3} eV^{2}$ for normal mass ordering (NO) and $2.20-2.54 \times 10^{-3} eV^{2}$ for inverted mass ordering (IO) \cite{valle}.
Table 1 presents the upper bounds on active-sterile mixing matrix elements and the experimentally allowed range of active-sterile mass-squared difference. Following are the 3$\sigma$ ranges of neutrino mixing matrix elements for active neutrinos:
\begin{eqnarray}
|U_{PMNS}|_{NO}\equiv \left(
\begin{array}{ccc}
 0.779- 0.842 & 0.52- 0.607 & 0.138- 0.161 \\
 0.205- 0.558 & 0.393- 0.716 & 0.618- 0.794 \\
 0.223- 0.568 & 0.417- 0.732 & 0.59- 0.772
\end{array}
\right),\\
|U_{PMNS}|_{IO} \equiv \left(
\begin{array}{ccc}
 0.779- 0.842 & 0.52- 0.607 & 0.140- 0.163 \\
 0.205- 0.556 & 0.394- 0.712 & 0.626- 0.792 \\
 0.227- 0.568 & 0.424- 0.732 & 0.592- 0.765
\end{array}
\right).
\end{eqnarray}
\begin{table}
\begin{center}
 \begin{tabular}{|c|c|c|}
   \hline
   Parameter & upper bound \\
   \hline
   $|U_{e4}|$  &  $< 0.228$ \ 95\% CL\\
   $|U_{\mu4}|$  &  $< 0.361$ \ 99\% CL\\
   $|U_{\tau4}|$  &  $< 0.548$ \ 99\% CL\\
   $\Delta m^{2}_{41}$ (eV$^2$) & 0.87 - 2.04\ \ \ 99.73\% CL\\
    \hline
 \end{tabular}
 \end{center}
 \caption{The current experimental bounds on sterile neutrino mixing parameters Ref.\cite{kopp} and mass-squared difference Ref. \cite{giunti}.}
\end{table}
In numerical analysis, we take the upper bound on sum of active neutrino masses $\Sigma m_{\nu}< 1 eV$. $\bar{\theta}_{14}$, $\bar{\theta}_{13}$ and $\bar{\theta}_{34}$ are free parameters which are varied randomly within the range [$0,\pi/2$]. The six neutrino mixing angles $\theta_{13}$, $\theta_{12}$, $\theta_{23}$ and $\theta_{14}$, $\theta_{24}$, $\theta_{34}$ are calculated using Eq.(10). We use Eq.(11) to calculate the neutrino mass eigenvalues $m_{1}$, $m_{2}$, $m_{3}$ and $m_{4}$. The unknown parameters $\bar{m}_{1}$ and $\bar{m}_{3}$ are generated randomly. The available experimental constraints on neutrino mass-squared differences and mixing matrix elements are used to restrict the unknown parameters. In Table 2, we have compiled the experimentally allowed ranges of various parameters of the model studied in the present work.
\begin{table}
\begin{center}
\begin{tabular}{|c|c|c|}
 \hline
 Parameter & Normal Mass Ordering (NO) & Inverted Mass Ordering (IO) \\
 \hline \hline
 $\bar{\theta}_{14}$  &  0 - 0.25 & 0 - 0.25 \\
 \hline
 $\bar{\theta}_{13}$  &  0.17 - 0.21 & 0.17 - 0.21\\
 \hline
 $\bar{\theta}_{34}$  &  0 - 0.35 & 0 - 0.35 \\
 \hline
 $|\bar{m}_{1}|$ (eV)  & 0 - 0.35 & 0.045 - 0.4 \\
 \hline
 $|\bar{m}_{2}|$ (eV) & 0.008 - 0.35 & 0.05 - 0.35 \\
 \hline
 $|\bar{m}_{3}|$ (eV) & 0 - 0.35 & 0 - 0.42 \\
 \hline
 $|f|$ (eV) & 0.035 - 0.42 & 0.023 - 0.42\\
 \hline
 $|g|$ (eV) & 0 - 0.35 & 0 - 0.3 \\
 \hline
 $|\bar{m}_{s}|$ (eV) & 0.8 - 1.5 & 0.8 - 1.5\\
 \hline \hline
 \end{tabular}
 \end{center}
 \caption{Experimentally allowed ranges of various parameters of the model.}
\end{table}

In our analysis, all the CP-violating phases are set to be zero and the effective Majorana mass $M_{ee}$ which determines the rate of neutrinoless double beta decay is given by
\begin{small}
\begin{equation}
M_{ee}=|m_{1} U_{e1}^{2}+m_{2} U_{e2}^{2}+m_{3} U_{e3}^{2}+m_{4} U_{e4}^{2}|.
\end{equation}
\end{small}
There are a large number of experiments such as CUORICINO \cite{cuor}, CUORE \cite{coore}, MAJORANA \cite{majo}, SuperNEMO \cite{super}, EXO \cite{exo} which aim to achieve a sensitivity upto 0.01 eV for $M_{ee}$.
The allowed ranges of $M_{ee}$ in our model for NO and IO are (0-0.35) eV and (0.015-0.4) eV, respectively.
\begin{figure}[H]
\begin{center}
\epsfig{file=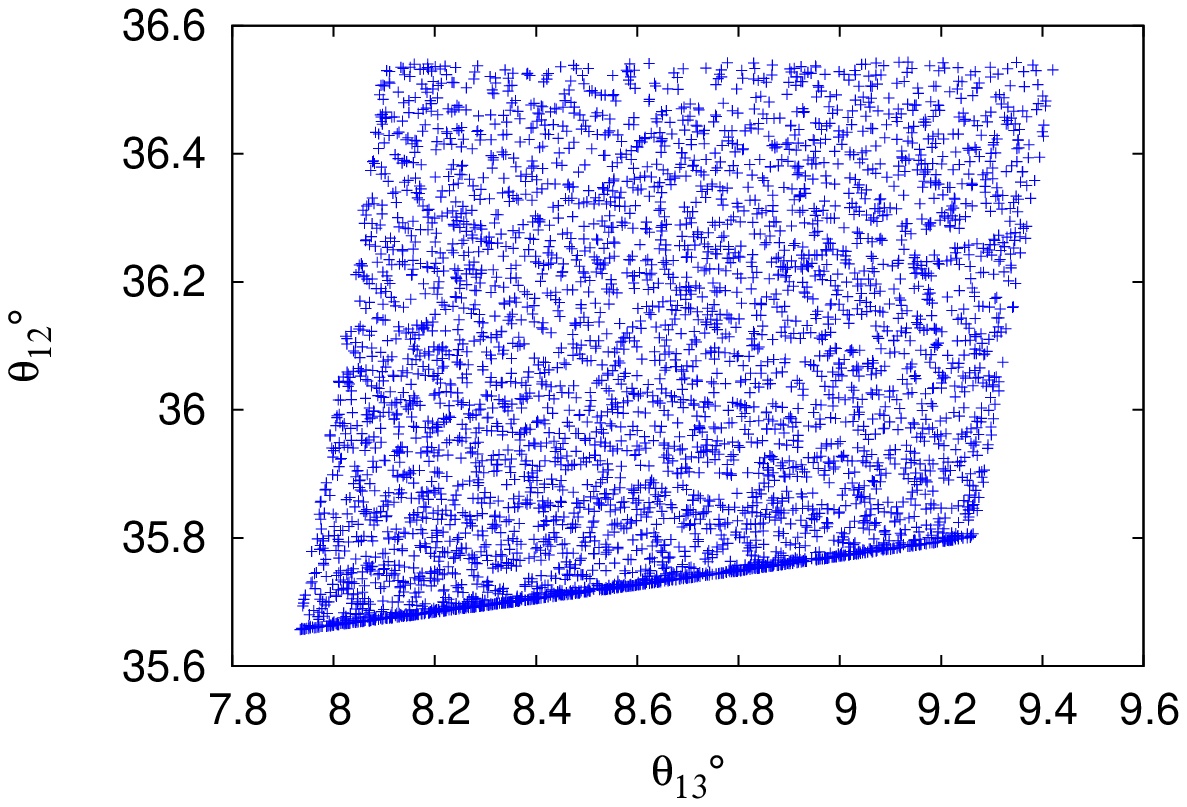, width=6cm, height=5cm}
\epsfig{file=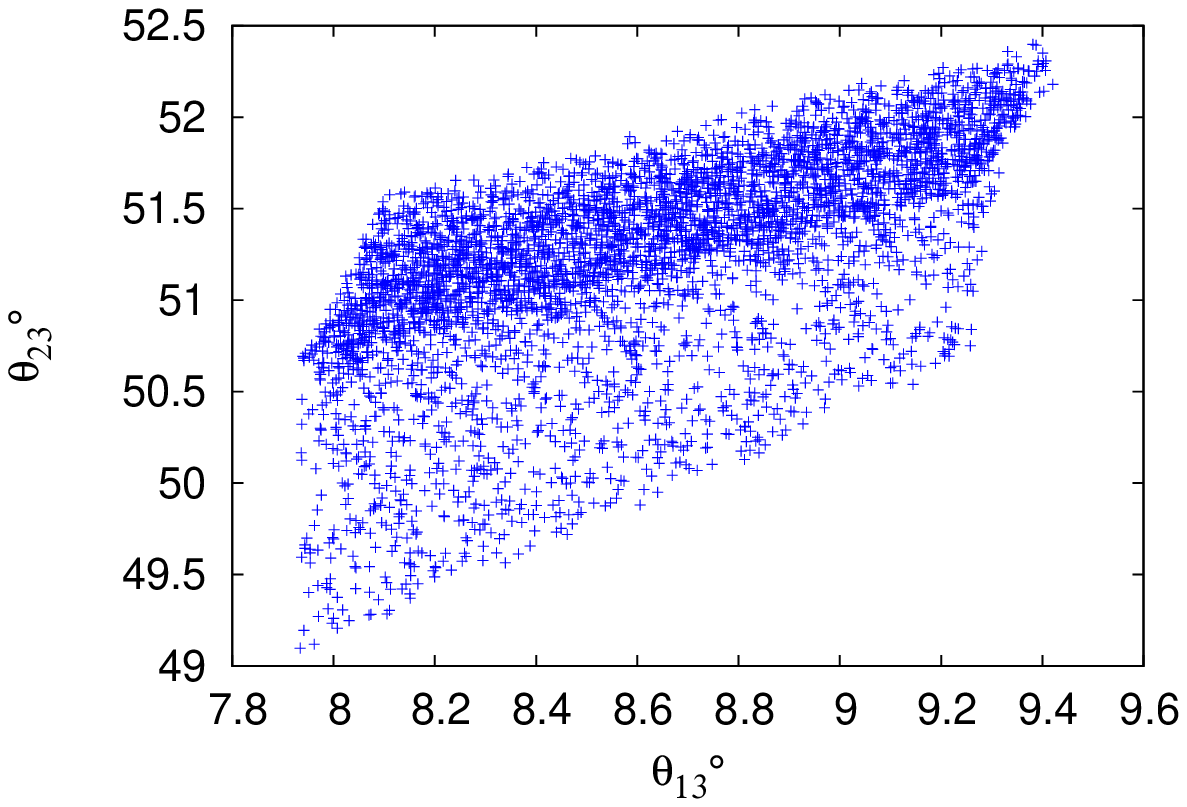, width=6cm, height=5cm}
\end{center}
\caption{Correlation plots among active mixing angles.}
\end{figure}
\begin{figure}[H]
\begin{center}
\epsfig{file=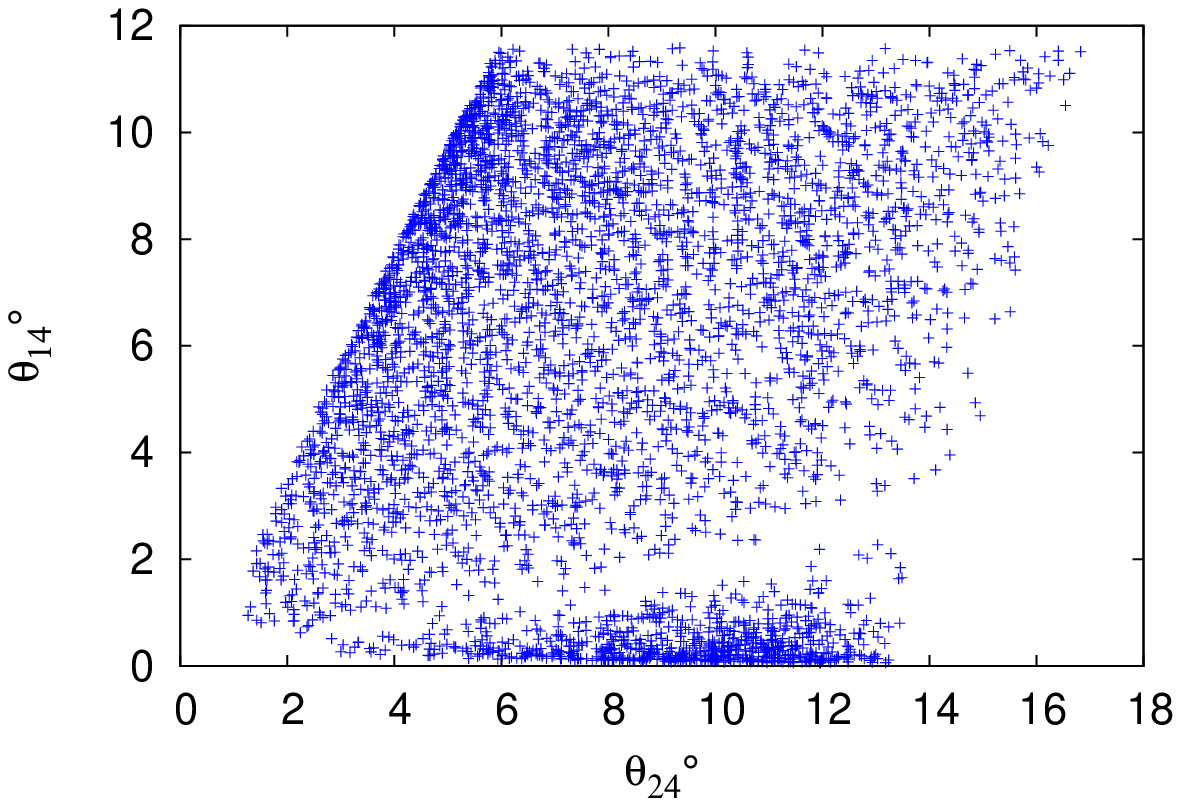, width=6cm, height=5cm}
\epsfig{file=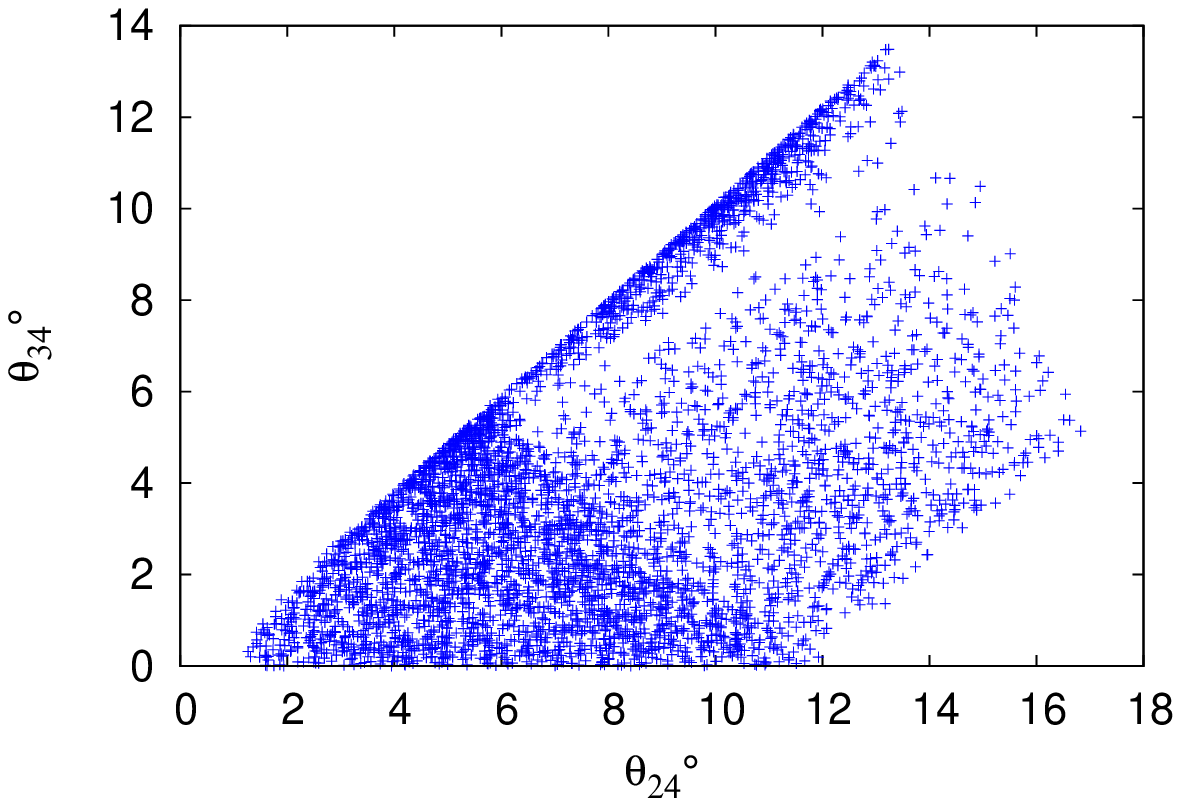, width=6cm, height=5cm}
\end{center}
\caption{Correlation plots among sterile mixing angles.}
\end{figure}
\begin{figure}[H]
\begin{center}
\epsfig{file=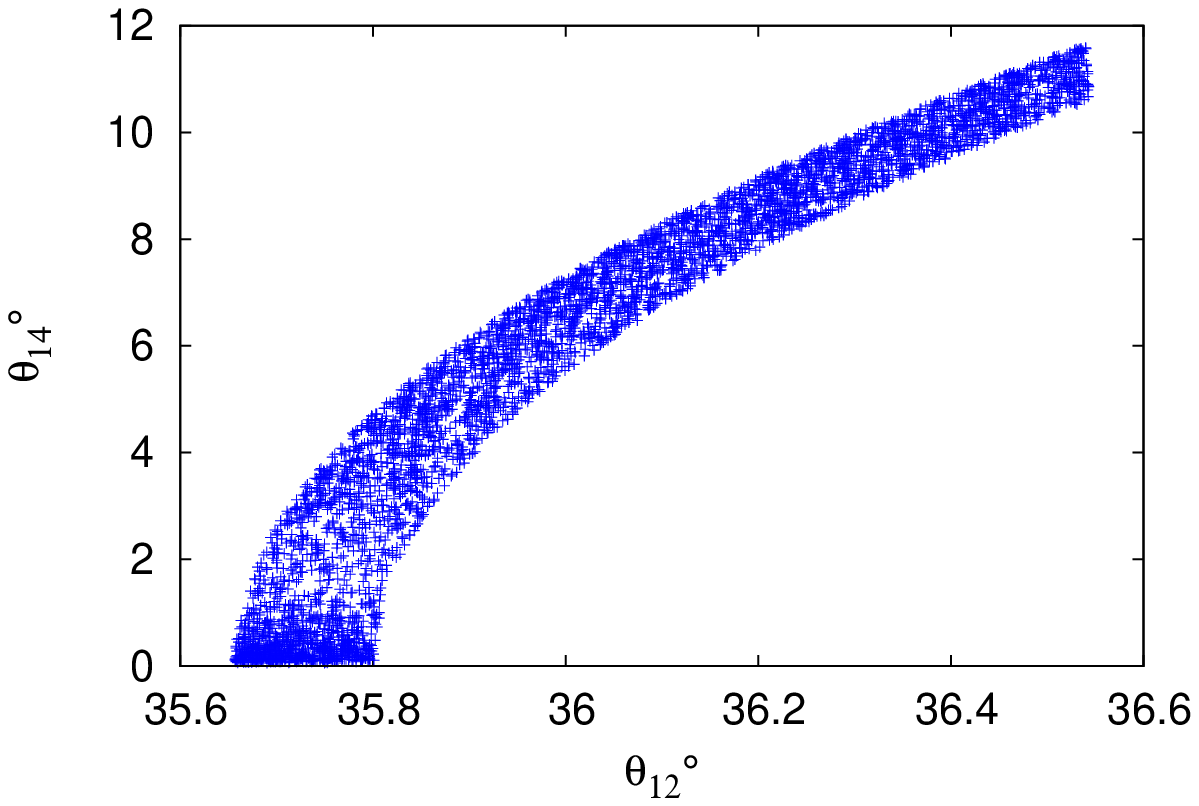, width=6cm, height=5cm}
\epsfig{file=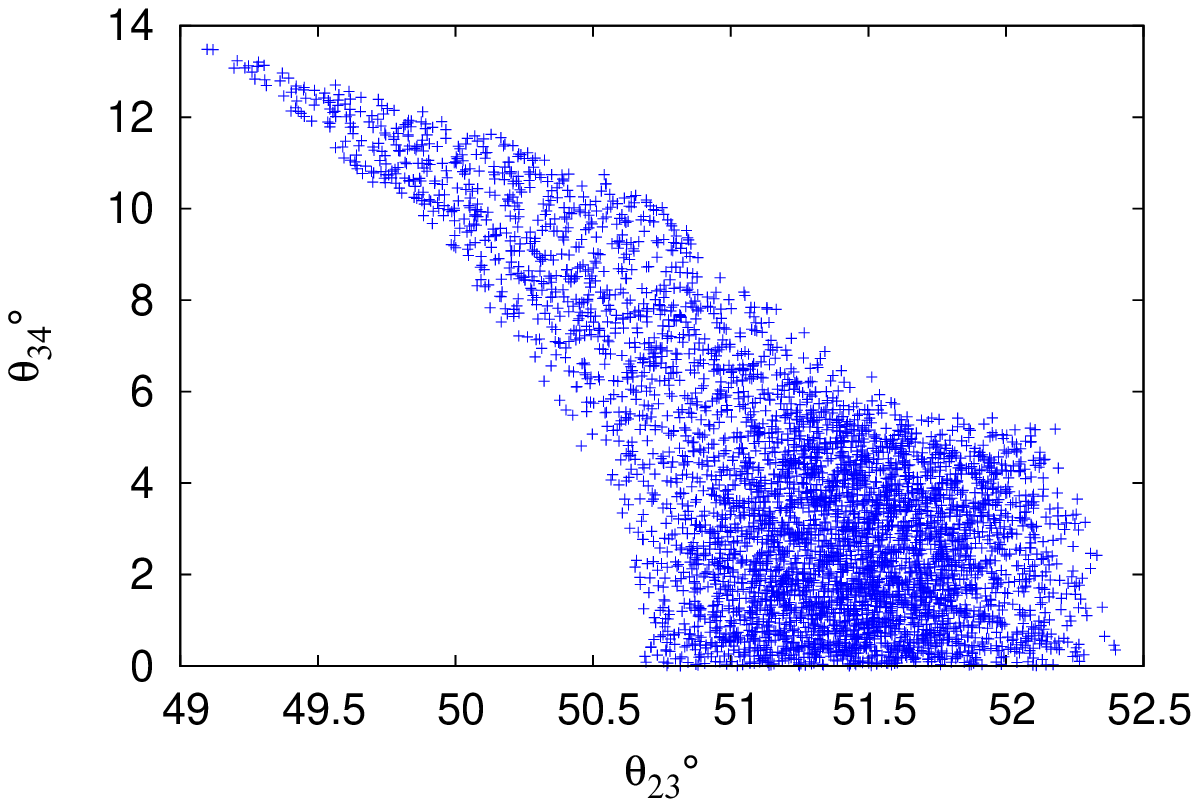, width=6cm, height=5cm}
\end{center}
\caption{Correlation plots between active and sterile mixing angles.}
\end{figure}

\begin{figure}[H]
\begin{center}
\epsfig{file=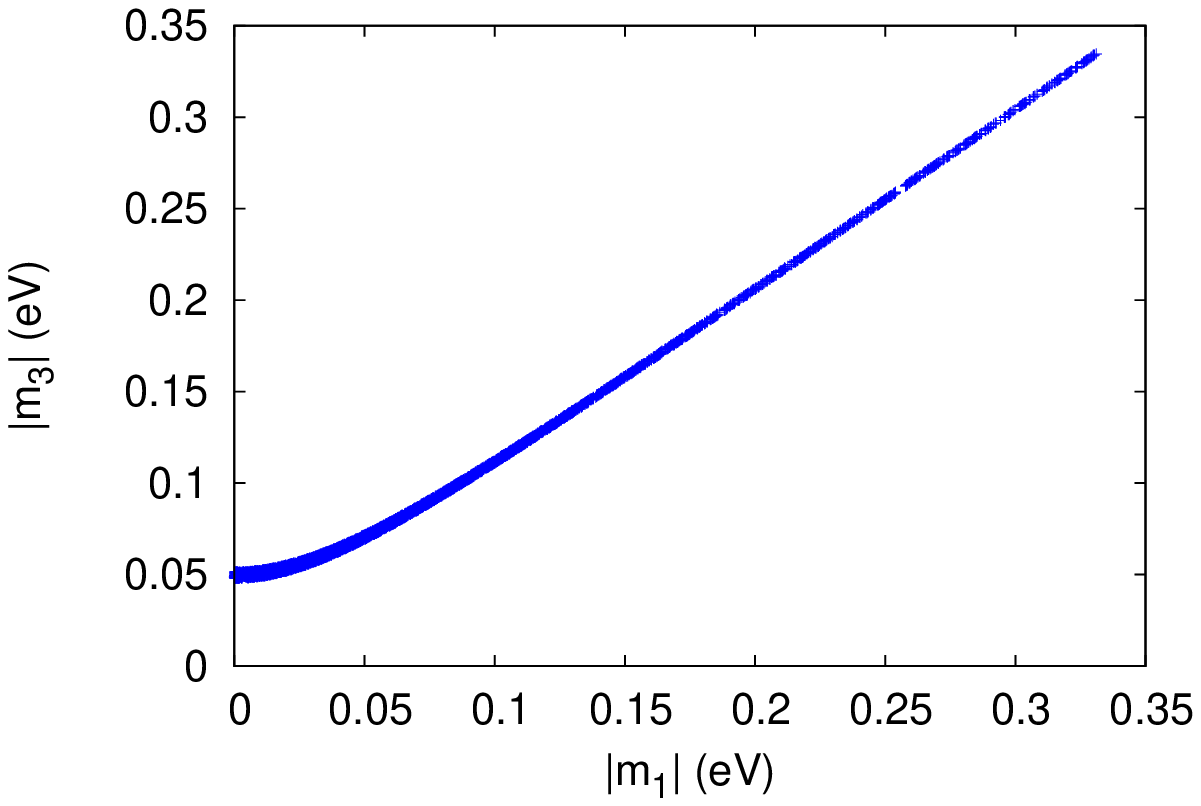, width=6cm, height=5cm}
\epsfig{file=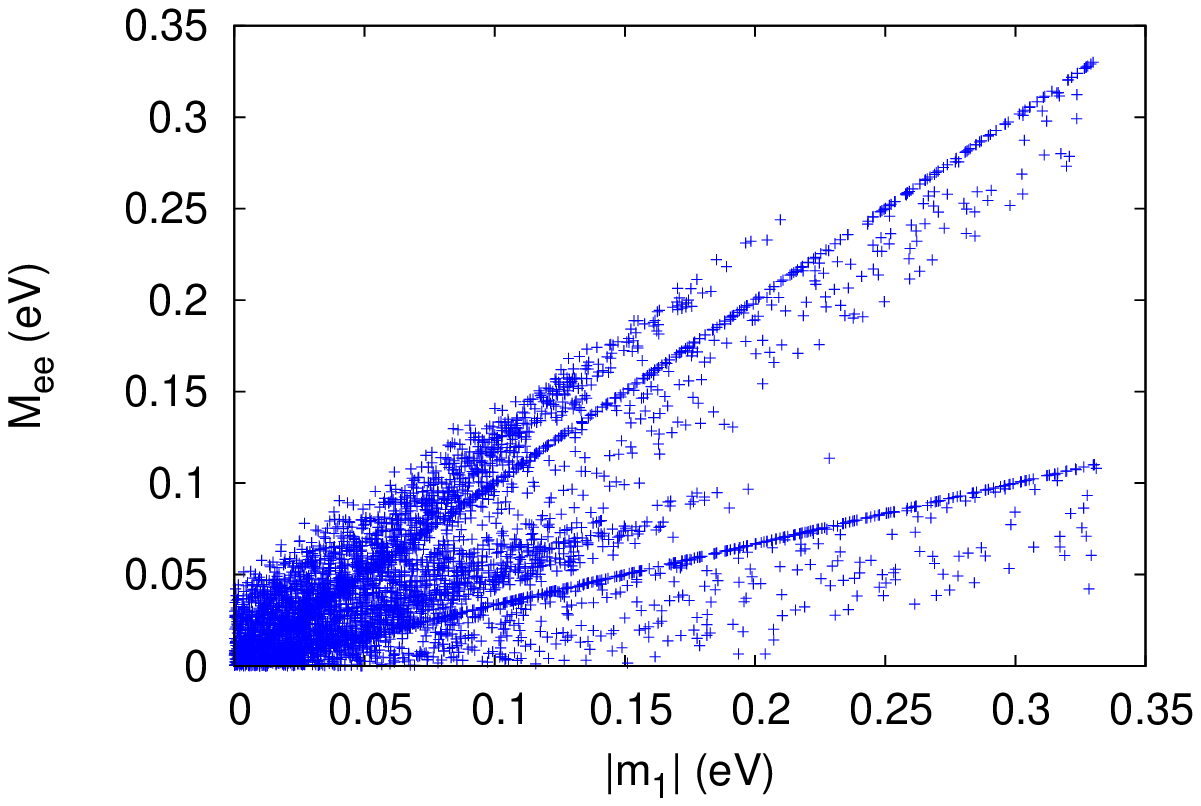, width=6cm, height=5cm}
\end{center}
\caption{Correlation plots for the normal mass ordering.}
\end{figure}
\begin{figure}[H]
\begin{center}
\epsfig{file=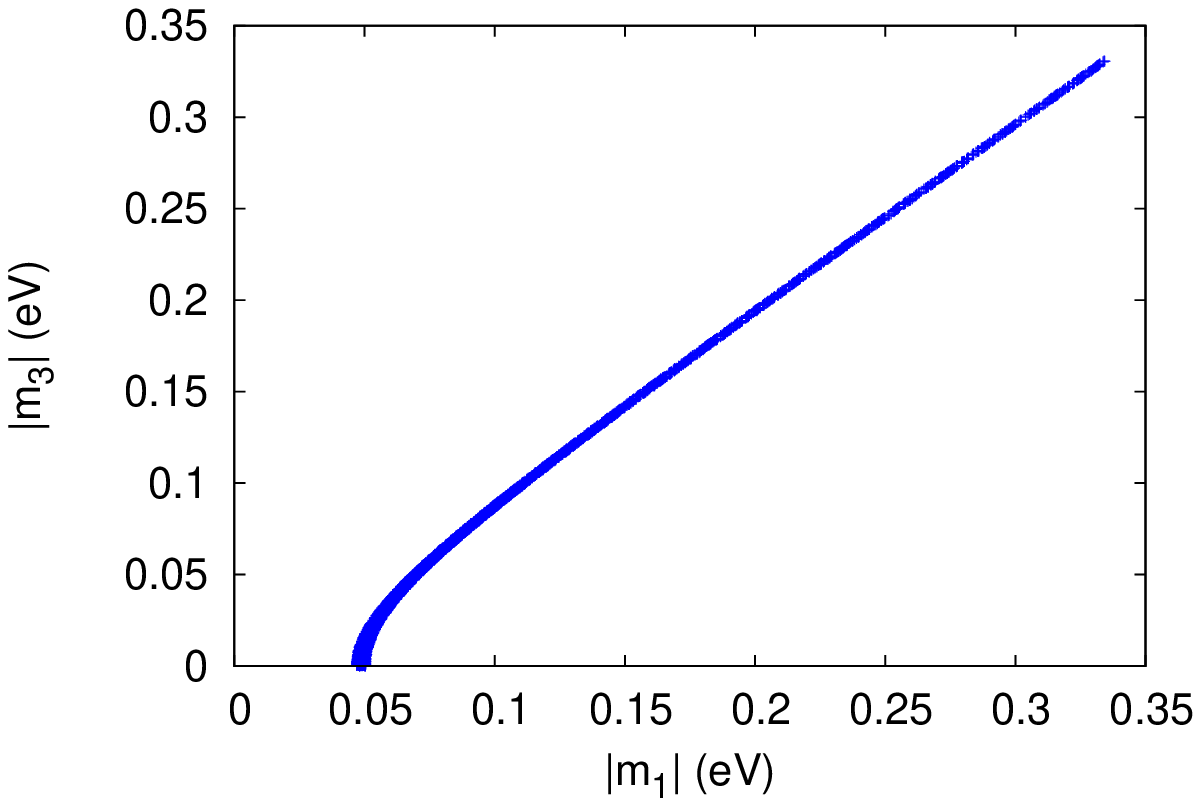, width=6cm, height=5cm}
\epsfig{file=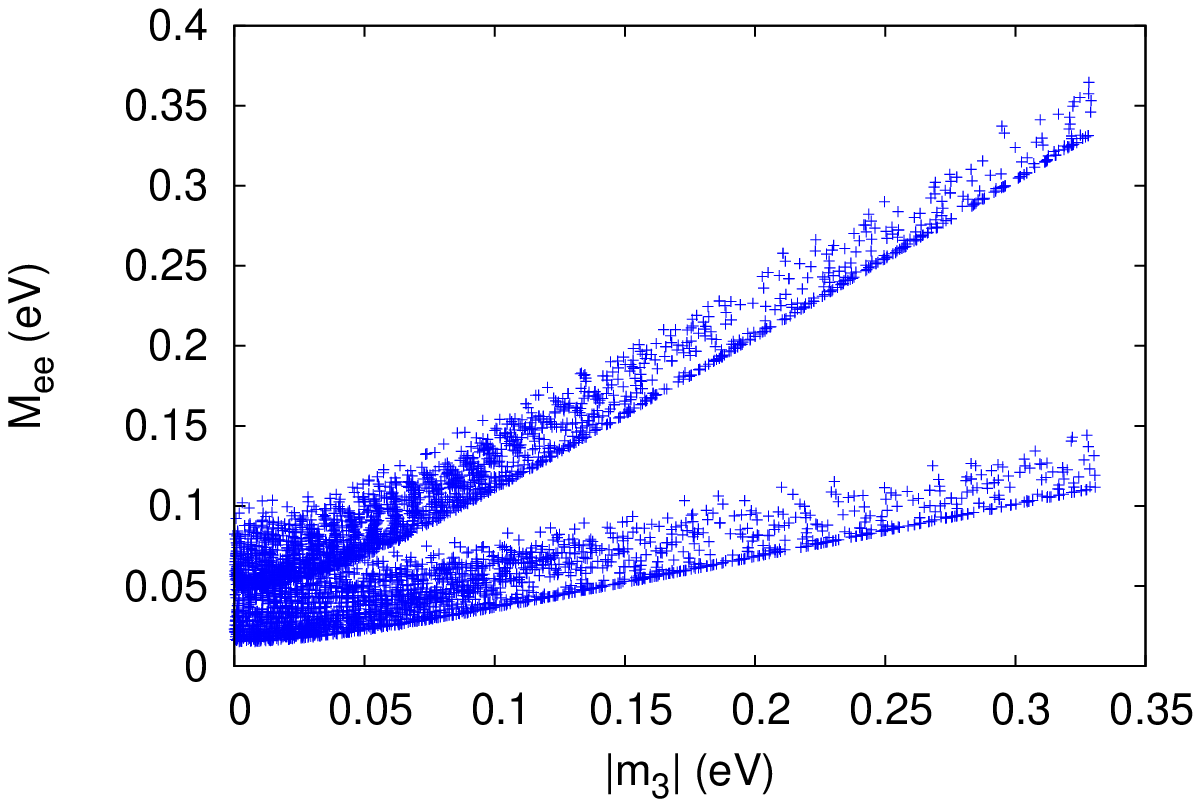, width=6cm, height=5cm}
\end{center}
\caption{Correlation plots for the inverted mass ordering.}
\end{figure}
Fig.(1) shows the correlations among active neutrino mixing angles. For the three neutrino case the correlation plot between $\theta_{12}$ and $\theta_{13}$ is a single line as shown in refs. \cite{carl,sanjeev,dev} but in the present case the plot is in the form of a band because of the presence of extra parameters coming from the sterile sector. The value of $\theta_{23}$ remains greater than $45^{\circ}$ in the present case. In fig.(2), we plot correlations between sterile angles.
The correlations between active and sterile mixing angles are shown in fig.(3). The correlations among neutrino mixing angles are the same for NO and IO. Only the mass matrix elements $\bar{m}_{1},\bar{m}_{2},\bar{m}_{3},f,g$ have different values for different mass orderings. Figs.(4) and (5) show plots for active neutrino masses $m_{1}$, $m_{3}$, and effective Majorana mass $M_{ee}$ for NO and IO, respectively.
\section{Summary}
In the present work, we have studied the phenomenological consequences of adding a light sterile neutrino to the active neutrinos. We examined the possibility of generating the necessary deviation from the TBM mixing by generating a non-zero $U_{e3}$ from active-sterile mixing. We have considered the simplest possible framework with only one sterile neutrino. The $3\times 3$ active neutrino sector of mass matrix has the TBM form. The presence of sterile neutrino and its mixing with active neutrinos leads to modification of the TBM pattern. The elements of the fourth row and the fourth column of the neutrino mass matrix can be chosen in such a way that the resulting neutrino mixing matrix has its second column coinciding with that of TBM. We found that a non-zero $U_{e3}$ within its experimental range can be successfully generated in this setting. Both normal and inverted mass orderings are allowed in this model. For simplicity, we have neglected CP violation in our analysis. The effective Majorana mass obtained in the present work lies well within the reach of forthcoming experiments. More stringent experimental constraints on sterile neutrinos can be obtained by the ongoing Daya Bay and upcoming JUNO experiments. In the present analysis, we have neglected the CP violation which otherwise may affect the analysis significantly. The analysis with the CP violating phases is, already, in progress.\\

\textbf{\textit{\Large{Acknowledgements}}}\\
The research work of R. R. G. is supported by the Department of Science and Technology, Government of India, under Grant No. SB/FTP/PS-128/2013. The research work of S. D. is supported by the Council for Scientific and Industrial Research, Government of India, New Delhi vide grant No. 03(1333)/15/EMR-II. S. D. gratefully acknowledges the kind hospitality provided by IUCAA, Pune.

\end{document}